\documentclass{article}

\usepackage{arxiv}

\usepackage[utf8]{inputenc} 
\usepackage[T1]{fontenc}    
\usepackage{hyperref}       
\usepackage{url}            
\usepackage{booktabs}       
\usepackage{amsfonts}       
\usepackage{nicefrac}       
\usepackage{microtype}      
\usepackage{lipsum}
\usepackage{graphicx}
\usepackage{amssymb}
\usepackage{amsmath,amsthm}
\usepackage{array}
\usepackage{subfig}
\usepackage{capt-of}
\usepackage{mathtools}
\usepackage[numbers, square, comma, sort&compress]{natbib}
\usepackage{enumitem}

\title{Impact of network assortativity on epidemic and vaccination behaviour}
\author{
	Sheryl L. ~Chang \textsuperscript{a}
	\thanks{Corresponding author} \\
	\texttt{ sheryl.chang@sydney.edu.au} \\
	\And
	Mahendra  Piraveenan \textsuperscript{a,b}\\
	\texttt{mahendrarajah.piraveenan@sydney.edu.au} \\  
	\And
	Mikhail Prokopenko \textsuperscript{a,c} \\
	\vspace*{1cm}
	\texttt{mikhail.prokopenko@sydney.edu.au} \\
	\textsuperscript{a}Complex Systems Research Group, Faculty of Engineering and IT \\
	The University of Sydney, NSW, 2006, Australia \\
	\textsuperscript{b}Charles Perkins Centre, The University of Sydney \\
	Johns Hopkins Drive, Camperdown, NSW, 2006, Australia\\
	\textsuperscript{c}Marie Bashir Institute for Infectious Diseases and Biosecurity \\
	The University of Sydney, Westmead, NSW, 2145, Australia\\
}
\begin{document}
	\maketitle
		\begin{abstract}
		The resurgence of measles is largely attributed to the decline in vaccine adoption and the increase in mobility. Although the vaccine for measles is readily available and highly successful, its current adoption is not adequate to prevent epidemics. Vaccine adoption is directly affected by individual vaccination decisions, and has a complex interplay with the spatial spread of disease shaped by an underlying mobility (travelling) network. In this paper, we model the travelling connectivity as a scale-free network, and investigate  dependencies between the network's assortativity and the resultant epidemic and vaccination dynamics. In doing so we extend an SIR-network model with game-theoretic components, capturing the imitation dynamics under a voluntary vaccination scheme. Our results show a correlation between the epidemic dynamics and the network's assortativity,  highlighting that networks with high assortativity tend to suppress epidemics under certain conditions. In highly assortative networks, the suppression is sustained producing an early convergence to equilibrium. In highly disassortative networks, however, the suppression effect diminishes over time due to scattering of non-vaccinating nodes, and frequent switching between the predominantly vaccinating and non-vaccinating phases of the dynamics.
		\end{abstract}

\section{Introduction}
\label{intro}
The World Health Organization (WHO) reported resurgence of measles across the globe in recent years \cite{measle_news}. For example, in Australia, New South Wales (NSW) Government issued 6 measles alerts in 2019 in which most infected cases were imported from overseas \cite{measle_NSW}. Health authorities confirmed that the rise of the highly infectious yet vaccine preventable disease is largely due to the decline in vaccination coverage and the increase in domestic and international travels. 

The high infectivity of measles is reflected in their high basic reproductive number ($R_0$), a measure defined as the average number of secondary cases arising from a typical infectious case in an otherwise susceptible population \cite{R0_beta,arino2003basic,miller2009spread}. Diseases with $R_0>1$ typically develop into an epidemic with a rapid onset characterised by an exponential increase in prevalence \cite{R0_ency}. The epidemic threshold (for example, $R_0=1$) is related to critical regimes and phase transitions which can be interpreted by methods of statistical physics \cite{wang2016statistical,harding2018,Harding2020}.    
The typical range of $R_0$ for measles is between 12 and 18, meaning that on average, each infectious individual with measles would infect 12 to 18 susceptible people \cite{measle_R0}. 

For  a highly infectious disease like measles, high vaccine adoption is extremely important for an epidemic suppression.  Vaccine adoption, driven by individual vaccination decisions, has been found very sensitive to many factors (e.g., cost of vaccination, disease prevalence, other individual behaviours, etc.), and can be characterised by highly nonlinear dynamics with oscillations \cite{bauch_2005_imi,imi_bauch_bha}. Various complex human interactions attribute to mobility and population mixing patterns, and  play a significant role in vaccine adoption. These non-trivial interactions are found to be particularly pronounced in scale-free networks \cite{Pastor-Satorras2}.

This study combines several fundamental me-thodologies, drawing from game theory and network theory. Such a multidisciplinary setting is required to address the key underlying challenges in computational epidemiology: time-dependent risks and varying imitation dynamics involved in vaccination decision-making, complex interaction and transmission patterns within the affected population, and nonlinear dependencies between the underlying network structure and the epidemic dynamics.

\textbf{\textit{(i) Imitation dynamics: game-theoretic modelling}}
The complex interplay between vaccination behaviour and the spatial spread of infectious diseases has attracted strong research interest \cite{vac_network,eksin2019control,chang2020game,soltanolkottabi2020game,Bauch_SR_2020}. The extent and dynamics of vaccine adoption emerge as a result of human interactions, driven by individual decision-making processes. In turn, this decision-making is based on individual evaluations of the benefits and weaknesses of specific strategies (i.e., to vaccinate or not to vaccinate). 
This process is often modelled by game theory where rational and self-centred individuals reach individual decisions after comparing their respective payoffs. Vaccination behaviour was shown to exhibit oscillatory dynamics when the individuals decide to vaccinate in response to current disease prevalence and social learning (i.e., imitating other people's behaviour), that is, when they are sufficiently responsive towards prevalence change \cite{bauch_2005_imi,imi_bauch_bha, Onofrio_2011,PHARAON201847}.

\textit{\textbf{(ii) Interaction patterns: network-theo-retic methods}}

A different challenge is that the the spatial spread of infectious diseases is dependent on the connectivity within a population,  with interactions only permitted between connected individuals~\cite{vac_network} or across different locations, in general. This feature is typically addressed by using network-theoretic, meta-population, or agent-based methods where interactions are only allowed between connected nodes, localities, or within specific mixing contexts~\cite{vac_network,ferguson2005strategies,germann2006mitigation,ajelli2010comparing,Acemod1,ABS,Acemod2,Zachreson2020,chang2020modelling,eksin2019control}. 
A network-theoretic component can be explicitly introduced to model spatial infection spread, using two approaches: 

(a) Local: person-to-person contacts at an individual level where each individual is modelled as a node and infection is only possible along edges \cite{network_game_bauch,sto_network_2010,sto_network_2013,sto_network_2012,sto_network_2012_memory,chao1,chao2,distancing_rs, Bauch_SR_2020}, and

(b) Global: meta-population level where each node represents a residential locality and people move across connected nodes \cite{patch,multi-city,SLC,Harding2020b}. 

Integrated with networks, recent models extend the game-theoretic framework to include imitation dynamics, where the imitation process is guided by the  network topology. The imitation process can take place at the local level where an individual imitates their neighbours~\cite{sto_network_2010,sto_network_2013,sto_network_2012,sto_network_2012_memory,chao1,chao2}, or at the global level by imitating the most successful strategy at a  residential locality, given some representation of travelling and population mixing~\cite{SLC}.

The topology of an underlying network plays a crucial role in the spreading process \cite{Pastor-Satorras1, Pastor-Satorras2}. Diverse topological impacts can be quantified by a multiplicity of network measures. In particular, assortativity was shown to correlate with the duration of an epidemic \cite{Newman2002},  maximum eigenvalue of the adjacency matrix was found to correlate with the epidemic threshold \cite{max_eigen1,max_eigen2}, and the maximum coreness was used to identify the key spreaders in a network \cite{core_epi}. 

In this paper, we focus on the role of assortativity in scale-free networks with the vaccine adoption resulting from human behaviour. In doing so, we adopt a game-theoretic framework with imitation dynamics. Our main contribution lies in untangling the interplay between the travelling patterns and the global vaccination and epidemic dynamics, providing a better understanding of the nonlinear dynamics which characterise both epidemics and vaccination behaviour. By setting up the travelling patterns as a scale-free network, we specifically investigate the role of the network's assortativity in affecting the resultant vaccination dynamics and suppressing the  epidemics.  Our model produces oscillatory dynamics for a range of assortativity (and disassortativity) values. In order to analyse the dynamics in relation to network robustness and stability of the dynamics, we trace the maximum eigenvalue of the adjacency matrix and the maximum coreness. As a result, we uncover the nonlinear correlation between epidemic dynamics and the assortativity in scale-free networks, and  highlight the important nonlinear effects of disassortative networks on suppressing epidemics.

The rest of this paper is structured as follows. Section \ref{material} describes the model (Section \ref{model}) and network configurations (Section \ref{metric}). Section \ref{results} shows the simulation results obtained for a range of scale-free networks. Finally, Section \ref{conclusion} summarises the findings. 

\section{Materials and methods}
\label{material}
\subsection{Model}
\label{model}
We adopt a model that considers the vaccination game with imitation dynamics where the individual's vaccination decision depends on the current prevalence, as well as on the vaccine adoption at their travelling destination \cite{SLC}. Importantly, the players of the game are parents: they decide, on behalf of their children, whether to vaccinate their child at birth. In doing so, the parents adopt either a vaccinator or a non-vaccinator strategy~\cite{bauch_2005_imi}. 

Consider a network with $M$ nodes, in which each node represents a suburb, $i \in V$, where  $i={1,2,\cdots M}$ \cite{multi-city, patch}. Travelling is only allowed between directly linked nodes $i$ and $j$, without hops on a daily basis, where a fraction of population living in node $i$ can travel to node $j$ and back, and vice-versa. The connectivity of suburbs and the fraction of people commuting between them are represented by the population flux matrix $\phi$. Entries in $\phi$ represent the fraction of population daily commuting from $i$ to $j$, $\phi_{ij} \in [0,1]$  (Equation \ref{phi}).  Some individuals may stay at their residential node (i.e. diagonal matrix $diag(\phi_{11},\phi_{22},...,\phi_{MM}) \ne 0$) and the population of each node is conserved so that each row in $\phi$ sums to 1.\\
\begin{equation}
\phi_{M \times M} =
\begin{bmatrix}
\phi_{11} & \phi_{12} & \cdots & \phi_{1M} \\
\phi_{21} & \phi_{22} & \cdots & \phi_{2M} \\
\vdots    & \vdots    & \ddots & \vdots \\
\phi_{M1} & \phi_{N2} & \cdots & \phi_{MM} 
\end{bmatrix}
\label{phi}
\end{equation}
To reach a vaccination (or non-vaccination) decision,  two factors need to be evaluated: the risk of infection based on current prevalence, and the most successful strategy based on social learning behaviour. At each time step, non-vaccinators evaluate their payoffs by weighing on the risk of vaccination from morbidity ($f_v$) and risk of non-vaccination from infection ($f_{nv}$),  as follows:
\begin{equation}
\begin{aligned}
f_v    &=-r_v \\
f_{nv} &=-r_{nv}mI(t) \\
\end{aligned}
\end{equation}
where $r_v$ is the morbidity from vaccination, $r_{nv}$ is the morbidity from infection, $m$ is the individuals' sensitivity to prevalence, and $I(t)$ is the current disease prevalence in the population fraction at time $t$.

To make a switch to the vaccinating strategy, the payoff gain must be positive $f_v-f_{nv}>0$. Individuals may also switch to the vaccinating strategy by imitating others (i.e., `imitation dynamics'), provided that vaccination is the most successful strategy in the population.

Let $x$ denote the relative proportion of individuals who choose to vaccinate, and assume that 
individuals use the combined imitation rate $\delta$ to sample and imitate strategies of other individuals. 
Because vaccination is an irreversible process, a vaccinated child cannot revert back to the non-vaccinated status. 
We note that $x$ is the proportion of individuals who choose to vaccinate at a given time (vaccinators), rather than the proportion of vaccinated children. Then the time evolution of $x$  can be defined as:
\begin{equation}
\begin{aligned}
\dot{x} & = \delta (1-x)x [-r_v+r_{nv}mI]\\
& = \kappa x(1-x)(-1+\omega I) \\
\end{aligned}
\label{x_bauch}
\end{equation}
where $\kappa=\delta r_v$ and $\omega=mr_{nv}/r_v$ \cite{bauch_2005_imi}.

A non-vaccinating individual (non-vaccinator) could imitate the strategy of a vaccinating individual (vaccinator)  if $-1+\omega I>0$. Conversely, if the risk of infection is not sufficiently high, individuals may choose the strategy of non-vaccination  ($-1+\omega I<0$). It is also assumed that vaccination is only provided to susceptible newborns with life-long protection against measles, meaning that individuals will not need to re-vaccinate.

Since both imitation and the current disease prevalence are based on the individual's travelling pattern, Equation \ref{x_bauch} is extended in a network setting. For any node $i \in V$, let $x_i$ denote the fraction of vaccinators in suburb $i$. Each day, non-vaccinators $(1-x_i)$ travel to suburb $j$ and encounter vaccinators  from node $k$. 

Every time a non-vaccinator  from $i$ comes in contact with a vaccinator from $k$, the imitation of the `vaccinate' strategy takes place with a combined imitation rate ($\delta$, i.e., $\kappa=\delta r_v$) applied to the difference in payoffs, $-1+\omega I_j>0$. For non-vaccinators, the switch to the vaccination strategy ($T^{NV \rightarrow V}$) is described as follows:
\begin{equation}
T_i^{NV \rightarrow V}=\kappa (1-x_i) \sum_{j=1}^{M} \sum_{k=1}^{M} \phi_{ij} \Theta (-1+\omega I_j) \phi_{kj} x_k
\end{equation}

Conversely, every time a vaccinator from $i$ comes in contact with a non-vaccinator from $k$, the imitation of the `not to vaccinate' strategy takes place with a combined imitation rate ($\delta$, i.e., $\kappa=\delta r_v$) applied to the difference in payoffs, $-1+\omega I_j<0$. For vaccinators, the switch to the non-vaccination strategy ($T^{NV \rightarrow V}$) is described as follows:
\begin{equation}
T_i^{V \rightarrow NV}=\kappa x_i \sum_{j=1}^{M} \sum_{k=1}^{M}  \phi_{ij} \Theta (-1+\omega I_j) \phi_{kj} (1-x_k) 
\end{equation}

The combined rate of change of the proportion of vaccinators in $i$ over time (i.e., the daily increase in $x_i$) can then be expressed by:

\begin{equation}
\begin{aligned}
\dot{x_i}  = &T_i^{NV \rightarrow V}+T_i^{V \rightarrow NV} \\
= & \kappa (1-x_i) \sum_{j=1}^{M} \sum_{k=1}^{M}  \phi_{ij} \Theta (-1+\omega I_j) \phi_{kj} x_k + \\
& \kappa x_i \sum_{j=1}^{M} \sum_{k=1}^{M} \phi_{ij} \Theta (-1+\omega I_j) \phi_{kj} (1-x_k)  \\
\end{aligned}
\end{equation}

Compared to the model presented in \cite{SLC}, the strategy switch is broken down into two separate processes: from non-vaccinators to vaccinators ($T_i^{NV \rightarrow V}$), and from vaccinators to non-vacci-nators ($T_i^{V \rightarrow NV}$). The strategy switch is governed by the Heaviside function $\Theta$ which ensures that only one process can occur at one time. 

The vaccination dynamics is then coupled with a standard SIR model where the population  at each node is categorised as susceptible ($S$), infected ($I$), and recovered ($R$). Successfully vaccinated newborns are directly transferred to the recovered class. Within each suburb (i.e., node), the population is homogeneous and conserved over time. In essence, the model divides the population into many homogeneous groups \cite{R0_beta}, based on their residential suburbs. The model is given by:  
\begin{equation}
\begin{aligned}
\dot{S_i} &=\mu(1-x_i)-\sum\limits_{j=1}^{M} \sum\limits_{k=1}^{M} \beta_j \phi_{ij} \frac{\phi_{kj}I_k}{\epsilon ^p_j} S_i-\mu S_i \\
\dot{I_i} &=\sum\limits_{j=1}^M \sum\limits_{k=1}^{M} \beta_j \phi_{ij} \frac{\phi_{kj}I_k}{\epsilon ^p_j}S_i-\gamma I_i -\mu I_i \\
\dot{R_i} &=\mu x_i+\gamma I_i-\mu R_i \\
\dot{x_i} &=\kappa (1-x_i) \sum_{j=1}^{M} \sum_{k=1}^{M}  \phi_{ij} \Theta(-1+\omega I_j) \phi_{kj} x_k + \\
&   \kappa x_i \sum_{j=1}^{M} \sum_{k=1}^{M}\phi_{ij}\Theta  (-1+\omega I_j) \phi_{kj} (1-x_k)  \\
\end{aligned}
\label{system2.1}
\end{equation}  
where  $\epsilon_l^p$ is a normalisation factor as the ratio between present population $N_j^p$ and the residential population $N_j$, and $\epsilon_j^p =\frac{N_j^p}{N_j} =\frac{\sum_{i=1}^{M} \phi_{ij}N_i}{N_j}.$ 

The dynamics of epidemic and vaccine adoption at each node is computed individually, and the global epidemic dynamics of the entire network can  be obtained by summing over all nodes.

\subsection{Network properties}
\label{metric}

Scale-free networks follow a power-law degree distribution with a few highly connected nodes (i.e., hubs) and numerous small-degree nodes. Such a tendency is commonly observed in real-world networks (e.g., air-traffic network, actor network, and the World Wide Web (WWW), etc.). Formally, the power-law distribution is defined as \cite{scale_free}:
\begin{equation}
P(k)=Ak^{-\gamma}u(k/N_p)
\label{scale_free}
\end{equation}
where $u$ is a step function specifying a cut-off at $k=N_p$. 

Assortative mixing, measured by assortativity coefficient, $r$ $(-1 \leq r \leq 1)$, is a preference for network nodes to connect to similar nodes where similarity can be defined in many ways, for example, in terms of node degree \cite{scale_free,Newman2002}.  For example, if highly connected nodes tend to connect to each other (e.g., actor network), the network is assortative ($r>0$); if highly connected nodes tend to link to small-degree nodes instead (e.g., the protein-interaction network of yeast), the network is disassortative ($r<0$) \cite{shannon,ass_bio,pira_addendum}. Our focus is to investigate the correlation between the assortativity and the epidemic severity. To do so, we will vary the assortativity of a scale-free network while preserving its scale-free properties. Xulvi-Brunet and Sokolov algorithm is used to rewire a scale-free network to a desired level of assortativity \cite{Xulvi-Brunet-Sokolov}  (see Appendix A for more details on rewiring algorithm). 

A $k{\text -}core$ is a maximal sub-network such that its nodes have at least $k$ degrees \cite{core_ori,network_pajek}. $k{\text -}core$ is obtained by recursively deleting all nodes of degrees less than $k$ and their edges and the remaining network is the $k{\text -}core$. This process is called $k{\text -}shell$ decomposition \cite{batagelj2003algorithm}. We are interested in the maximum $k{\text -}core$ (i.e., maximum coreness), $k_s$, which characterises the nodes located in the innermost shell.  In assortative networks, many of these nodes located at the core of the network are hubs and have been shown to be ``influential spreaders"  for forming a reservoir to sustain epidemic spreading \cite{core_epi,Newman2002}, prolonging the duration of an epidemic outbreak.  In disassortative networks, however, hubs are located at the periphery of the network and tend to be connected to peripheral nodes, suppressing epidemic spread and resulting in shorter epidemic periods \cite{core_epi}. It is also found that assortative rewiring increases $k_s$ by creating more `shells' (see Appendix A Figure \ref{rewire_k_s} for network visualisation).

\begin{figure}
	\centering
	\includegraphics[width=0.5\textwidth,trim={1cm 0cm 0cm 0cm},clip]{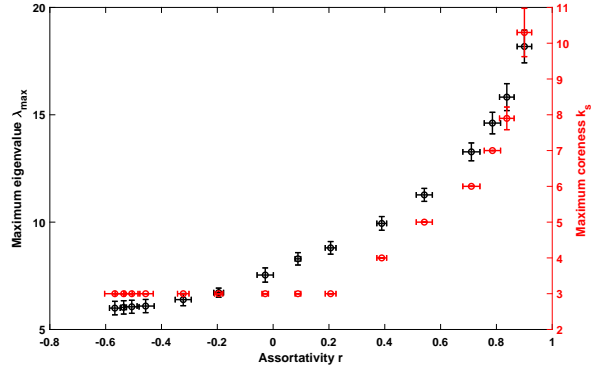}
	\caption{Scale-free networks: relationship between assortativity, the maximum eigenvalue and the maximum coreness. Y-axis (left, in black): maximum eigenvalue $\lambda_{max}$; Y-axis (right, in red): maximum coreness, $k_s$. Each data point is averaged over 10 runs. Error bars denoted standard deviation.}
	\label{r_eig}
\end{figure}

The maximum eigenvalue of the adjacency matrix, $\lambda_{max}$, has been related to the network's robustness during diffusion processes (e.g., virus propagation) \cite{robustness}. Importantly, $\lambda_{max}$ is in an inverse relationship with the epidemic threshold ($\tau$) for arbitrary graphs ($\tau=1/\lambda_{max}$), indicating that networks with higher $\lambda_{max}$ will have lower epidemic threshold. Therefore, these networks are more vulnerable to show an epidemic outbreak after an initial infection \cite{max_eigen1,max_eigen2}. The maximum eigenvalue and the maximum coreness are found to increase with assortativity. We illustrate the dependency in Figure \ref{r_eig}.

In the absence of vaccine, highly assortative networks are found to have high $k_s$ and high $\lambda_{max}$ and may be subject to earlier outbreak  \cite{max_eigen1,max_eigen2} with prolonged epidemic period \cite{core_epi,Newman2002}, while disassortative networks are less vulnerable to epidemic. With the addition of voluntary vaccination, one may make a conjecture that the population in highly assortative networks can  opt to vaccinate if the epidemic is sustained, which will consequently result in reducing the duration of epidemic. A similar question is whether, in presence of vaccination, the highly disassortative networks continue to be better at suppressing epidemics. We verify these conjectures by simulating epidemic spread across a range of scale-free networks, varying their assortativity while preserving the degree distribution. 

We set the epidemiological parameters to a scenario motivated by measles epidemics ($R_0=15$ \cite{measle_R0}), while assuming uniform initial conditions and the population's responsiveness to a prevalence change across all nodes. A small fraction of the initial infected population is deployed in all nodes to evaluate the impact of the travelling diffusion. We refer to Table \ref{parameter} in Appendix B for all parameters used.

\section{Simulation results}
\label{results}
We first analyse how assortativity affects the severity of the epidemics and the vaccination dynamics, and then interpret the results using three different network measures: assortativity $r$, the maximum eigenvalue $\lambda_{max}$, and the maximum coreness $k_s$. The severity of epidemics is quantified by the prevalence peak ($I$) and the cumulative incidence
(i.e., proportion of newly infected individuals in the population at risk) during the first outbreak where the maximum prevalence occurs. We refer to Appendix B for more details about the relationship between the disease prevalence and cumulative incidence.

\subsection{Oscillatory dynamics}
\begin{figure*}
	\includegraphics[width=\textwidth,trim={2.5cm 0cm 3cm 0cm},clip]{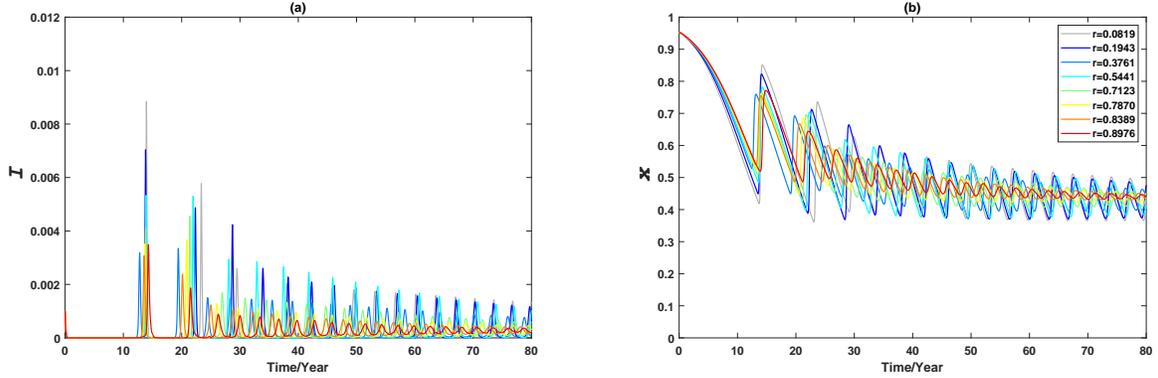}
	\caption{Epidemic dynamics of a set of scale-free networks ($N=3000, \gamma=2.75, k_0=1,\langle k \rangle \approx 4$), with assortative rewiring. Time series of (a) \textit{I}, disease prevalence, and (b) \textit{x}, fraction of vaccinators. Initial conditions: $I_0=0.001, S_0=0.05,x_0=0.95$. Behaviour parameters: $\omega=3500, \kappa=0.001$} 
	\label{osci_ass_3500}
\end{figure*}

\begin{figure*}
	\includegraphics[width=\textwidth,trim={2.5cm 0cm 3cm 0cm},clip]{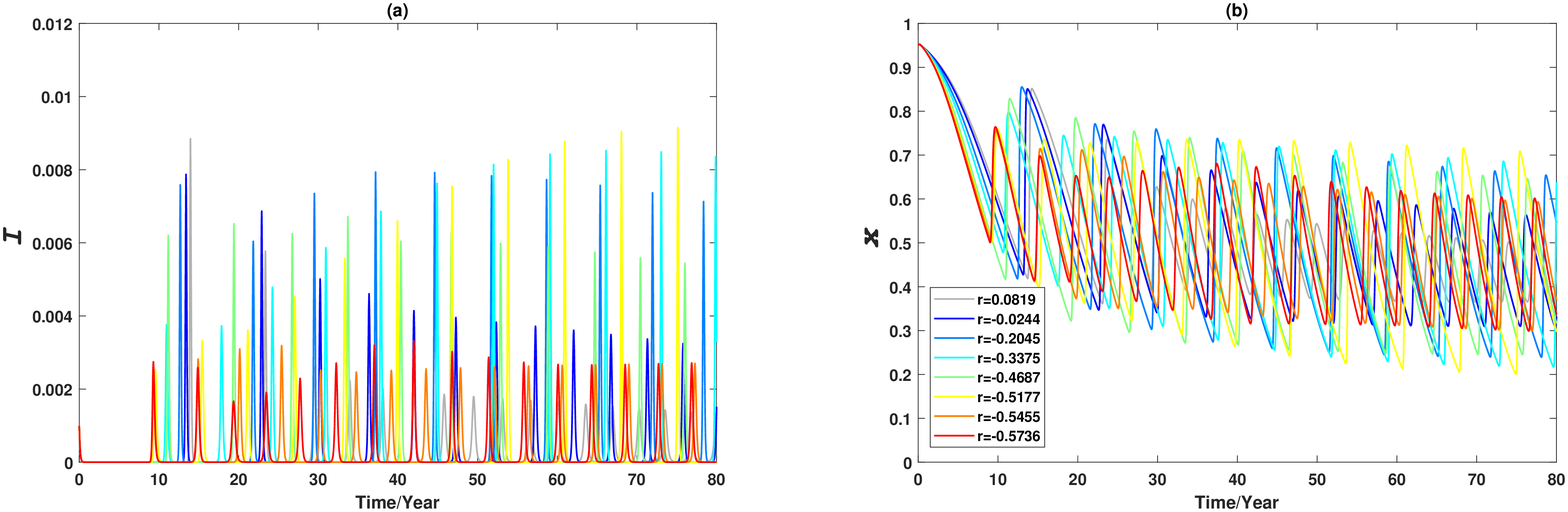}
	\caption{Epidemic dynamics of a set of scale-free networks ($N=3000, \gamma=2.75, k_0=1,\langle k \rangle \approx 4$), with disassortative rewiring. Time series of (a) \textit{I}, disease prevalence, and (b) \textit{x}, fraction of vaccinators. Initial conditions: $I_0=0.001, S_0=0.05,x_0=0.95$. Behaviour parameters: $\omega=3500, \kappa=0.001$} 
	\label{osci_dis_3500}
\end{figure*}

Figure \ref{osci_ass_3500} and Figure \ref{osci_dis_3500} show the oscillatory dynamics in disease prevalence ($I$) and vaccine adoption ($x$) for the networks across the simulated assortativity range ($-0.5671 \leq r \leq 0.9002$) at relatively high $\omega=3500$. The oscillations indicate that responsive individuals would react to changes in prevalence and consequently choose to vaccinate, however, the high level of vaccine adoption is not sustainable, as soon as the prevalence reduces (Figure \ref{osci_ass_3500} (b)), in concordance with previously established results~\cite{bauch_2005_imi, SLC}. As a result, the individuals would consequently choose the non-vaccination strategy because the risk of infection is low,  since there is no change in disease prevalence. Moreover, we find that both prevalence and vaccine adoption reach an early convergence to the mixed, endemic equilibrium in highly assortative cases (i.e., $r \gtrapprox 0.75$). On the other hand, oscillations are sustained in all studied disassortative networks (Figure \ref{osci_dis_3500} (b)). In addition, the first outbreaks in highly disassortative  networks occur noticeably earlier, albeit with a relatively low prevalence peak ($I_{max} \approx 0.0035$), as shown in Figure \ref{osci_dis_3500} (a). 

\begin{figure*}
	\includegraphics[width=\textwidth,trim={2.5cm 0cm 3cm 0cm},clip]{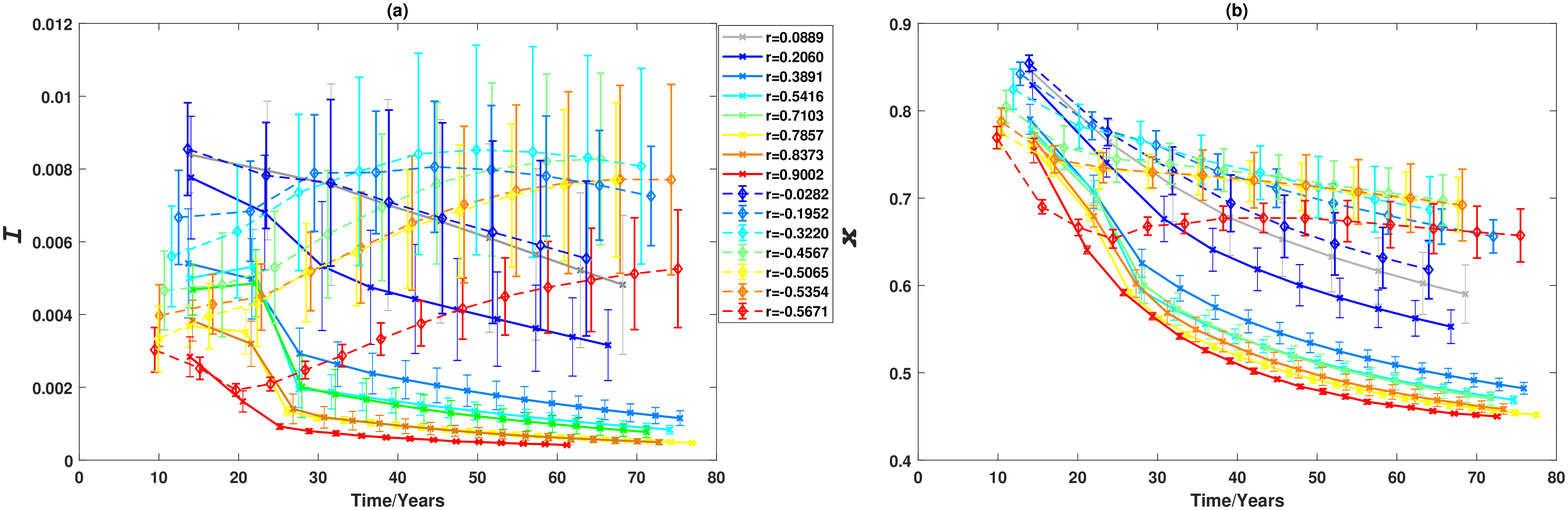}
	\caption{Epidemic dynamics of scale-free network ($N=3000, \gamma=2.75, k_0=1,\langle k \rangle \approx 4$), with assortative and disassortative rewiring. Time series of peak amplitude for (a) \textit{I}, disease prevalence, and (b) \textit{x}, fraction of vaccinators. Each data point is the average of 10 runs at the same setting. Error bars denote standard deviation. Solid line: assortative range; dashed line: disassortative range. Peaks amplitudes are identified if the differences between adjacent peaks are greater than threshold value $\hat{x}=0.01,\hat{I}=0.0002$. Initial conditions: $I_0=0.001, S_0=0.05,x_0=0.95$. Behaviour parameters: $\omega=3500, \kappa=0.001$.} 
	\label{assor_amp}
\end{figure*}

In order to account for network variability, each rewiring setting is then repeated for 10 scale-free networks with the same topological properties. For better readability, only prominent peaks in $x$ and $I$ are shown in Figure \ref{assor_amp}.  We observe that the prevalence peaks are correlated with the network's assortativity (or disassortativity). Figure \ref{assor_amp} shows that epidemics are generally better suppressed in networks with high positive $r$ (assortative), where the prevalence steadily diminishes over time, and this effect increases with $r$. The non-assortative networks (i.e., $|r| \lessapprox 0.2$) have a less beneficial effect than the assortative ones, but retain the suppression tendency over the years.  However, the long-term suppression effect which is sustained in both assortative and non-assortative networks, is disrupted in disassortative topologies. In these cases ($r \lessapprox -0.2$), we observe only short- to medium-term benefits. Specifically, the prevalence peaks start to grow after year 20, and this tendency intensifies over time. The suppression effect is somewhat regained for highly disassortative networks (e.g., $r \lessapprox -0.55$), but even in these cases, the effect is short-lived, and starts to diminish after the year 20. 

In order to explain these nonlinear effects, we explore the role of network topology by tracing the vaccine adoption over time in individual nodes. In doing so, we categorize  the network nodes into majority vaccinators or majority non-vaccinators, by applying a simple threshold ($\bar{x}=0.5$) to the fraction of vaccine adoption in a given node at each time step. More precisely, any node $i$ where the majority of population chooses the vaccination strategy ($x_{i}>\bar{x}$) is identified as a vaccinator node. Conversely, a node $i$ with the majority of population choosing the non-vaccination strategy ($x_{i}<\bar{x}$) is deemed to be a non-vaccinator node.    

Thus, any network is partitioned into two  sub-networks, formed by majority vaccinators or majority non-vaccinators only. Each majority sub-network is then traced over time in terms of the number of nodes (Figures \ref{node_ass} and \ref{node_dis}), and the size of the largest connected subgraph (LCS) (Figures \ref{LCS_ass} and \ref{LCS_dis}), as shown in Appendix C.   
There are two features we observe in the respective network dynamics. 

Firstly, there are alternating phase transitions in the connectivity of the majority sub-networks: at a critical threshold, there is a sudden formation of a giant connected subgraph emerging after a period of low connectivity within the respective majority sub-network. Strictly speaking, a phase transition is defined in the space of some control parameter, such as an epidemic threshold (typically, the underlying reproductive number $R_0$~\cite{artalejo2013exact,artalejo2014stochastic,Yagmur,Harding2020}), the per capita commuting rate \cite{balcan2011phase},  the rationality level adopted by decision-making individuals~\cite{Harding2020b}, and so on. A critical regime may also be traced over time if the control parameter changes concurrently. In the cases considered in our study, as the infection affects the vaccinating behaviour of the decision-making individuals, the resulting adoption of the vaccinating (or non-vaccinating) strategy spreads through the network, with each of the strategies dominating at alternating phases. 

It is evident, from the analysis of the number of nodes and the size of the LCS, that each phase is completely dominated by one of the strategies:  the maximal size of the dominating LCS, as well as the corresponding number of nodes, almost reach the size of the entire network. In other words, the giant subgraph of each sub-network encompasses almost all of the network nodes. The dynamics clearly display  an antiphase behaviour between the  high- and low-connectivity phases.  For example, when the majority non-vaccinator nodes dominate the network, the nodes adopting the opposing, vaccinating, strategy are mostly disconnected and their sub-network is fragmented. This fragmentation, reflected in the very low connectivity, will be rapidly replaced with a new giant subgraph (now with majority vaccinating nodes), formed at the next phase transition.  We also point out that the alternating phase transitions tend to switch with a higher frequency as the (dis)assortativity $|r|$ grows.

The second observed feature is that the nature of low connectivity within the fragmented sub-networks differs between the assortative and disassortative ranges.  By examining the minimal sizes of the LCS (Figure \ref{LCS_ass} and \ref{LCS_dis}), we note that the minimum LCS sizes in both the vaccinator and non-vaccinator sub-networks are significantly larger in assortative networks than in their disassortative counterparts. This observation suggests that vaccinator and non-vaccinator nodes are more clustered in assortative networks (forming ``pockets'' of the respective strategy adopters). Furthermore, during the period before the dynamics converge in highly assortative topologies ($r \gtrapprox 0.7$),  the minimum LCS sizes of the majority vaccinating sub-networks  are larger than those in the majority non-vaccinator sub-networks. This difference in sizes is not pronounced in the disassortative range, and we conclude that, during the low-connectivity phase, the vaccinator nodes are scattered, rather than clustered, in disassortative networks.  Arguably, the scattering of vaccinator nodes makes them particularly prone to a strategy switch in disassortative topologies.

In summary, there is a confluence of (i) the scattering of  vaccinator nodes during the low connectivity phase, and (ii) the alternating phase switching, at a higher frequency of phase transitions for higher $|r|$.  This combination makes the disassortative networks less conducive for maintaining a vaccinating strategy in the long-term, causing the epidemic suppression effect to diminish over time.

\subsection{Correlation between network measures and epidemic severity}
To further evaluate the severity of the epidemic, we investigate the relationship between assortativity and the cumulative incidence under the initial vaccine adoption $x_0=0.95$ and $S_0 = 0.05$, i.e. a near herd immunity scenario where the vast majority of population is protected. 

\begin{figure*}
	\includegraphics[width=\textwidth,trim={3.5cm 0.5cm 3.5cm 0.5cm},clip]{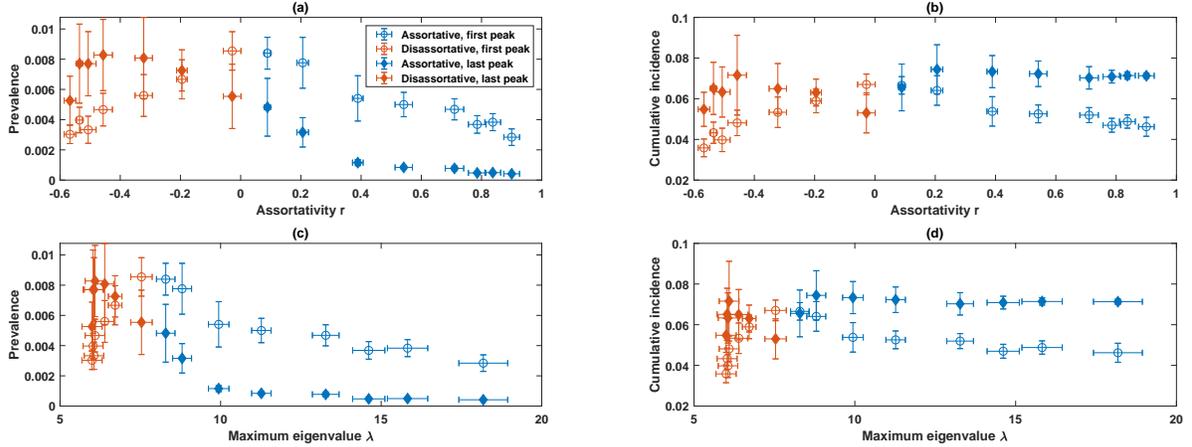}
	\caption{Relationship between epidemic dynamics and network properties. Epidemic dynamics is evaluated by the prevalence and cumulative incidence of the first (circle, unfilled) and the last wave (diamond, filled). Note that in highly assortative network, detection of last wave occurs noticeably earlier than in non-assortative network. For consistency, cumulative incidence of the last wave is evaluated between year 75 to year 80. Network properties are evaluated by assortativity ($r$) and the maximum eigenvalue, $\lambda$. Data is extracted from Figure \ref{r_eig} and Figure \ref{assor_amp}. Error bars denote standard deviation over 10 runs.} 
	\label{SF_ass}
\end{figure*}

Figure \ref{SF_ass} shows the dependency between network properties (assortativity and the maximum eigenvalue) and epidemic severity.  Two cases are presented: the first and the most significant epidemic wave (circle, unfilled), and the last wave towards the end of simulation time (diamond, filled). 

Examining Figure \ref{SF_ass} (a) and (b), we observe that during the first wave of epidemic, the population is at lower risk in both disassortative networks (i.e., $r \lessapprox -0.45$) and assortative networks (i.e., $r \gtrapprox 0.4$). In contrast, the population is at the highest risk (i.e., having the highest cumulative incidence) in non-assortative networks (i.e., $|r| \lessapprox 0.2$). When the epidemic approaches the end of simulation time, as $r$ increases, the amplitude of prevalence peaks continues to reduce in highly assortative networks as the epidemics converge to the mixed endemic equilibrium. This represents a ``flattened curve" which is generally a preferred outcome that leads to a reduced load on health care resources at a given time (Figure \ref{SF_ass} (b)).

\begin{figure*}
	\centering
	\includegraphics[width=\textwidth,trim={3cm 0cm 3cm 0cm},clip]{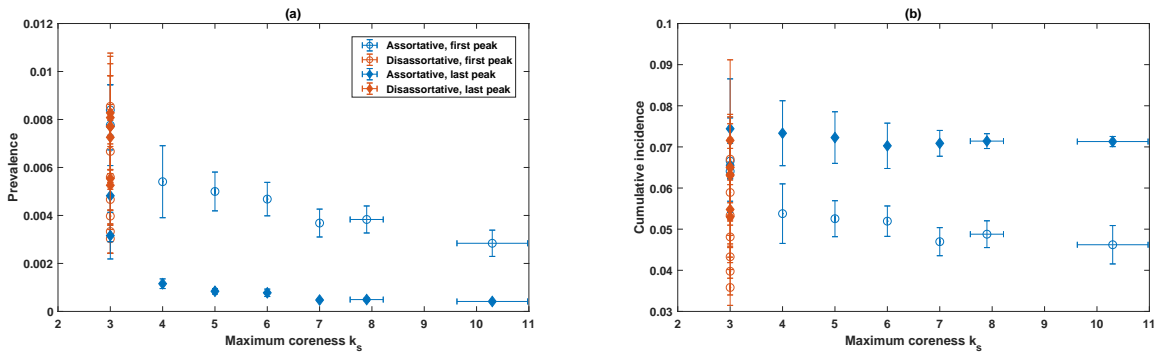}
	\caption{Relationship between the maximum coreness $k_s$ and epidemic severity. Epidemic dynamics is evaluated by the prevalence and cumulative incidence of the first (circle, unfilled) and the last wave (diamond, filled). Each data point is averaged over 10 runs. Error bar denotes standard deviation.} 
	\label{epi_k}
\end{figure*}

Similar dependency is also observed in terms of the maximum eigenvalue of the adjacency matrix, $\lambda_{max}$ (Figure \ref{SF_ass} (c) and (d)). We find that both $r$ and $\lambda_{max}$ display a profile where the cumulative incidence is lower in assortative and disassortative networks at the start of the epidemic. Although a similar dependency is also observed in the relationship between epidemic severity and the maximum coreness $k_s$, the latter does not adequately differentiate between the highly disassortative networks, as $k_s$ only increases with the assortative rewiring, and is otherwise identical across disassortative and non-assortative networks (Appendix C, Figure \ref{epi_k}). 

\section{Conclusion}
\label{conclusion}

In this work, we investigated how assortativity affects epidemic and vaccination dynamics in scale-free networks by adopting a game-theoretical SIR-network model with imitation dynamics. We first illustrated the dependencies between assortativity and other network-theoretic measures, i.e., maximum coreness ($k_s$) and maximum eigenvalue of adjacency matrix ($\lambda_{max}$). While $\lambda_{max}$ increases monotonically with assortativity for the tested assortativity range ($-1 \leq r \leq 1$), $k_s$ only increases with assortative rewiring.  

Previous findings suggested the epidemics are more likely to sustain in highly assortative networks. By introducing voluntary vaccination, it is expected that the epidemic duration may be reduced, as some individuals living in the `hub' nodes may choose to vaccinate.  This conjecture was verified by  simulations of the epidemic spread on scale-free networks with varying assortativity. 

We observed oscillatory dynamics in both disease prevalence ($I$) and vaccine adoption ($x$),  concurring with earlier studies~\cite{bauch_2005_imi,samit_bauch_2011_delay}.  In assortative networks, when $I$ converges to the mixed, endemic equilibrium, the convergence of prevalence also leads to an equilibrium in the vaccine adoption, typically under 50\%. We find that highly assortative networks have a stronger sustained effect in suppressing epidemics, while such a benefit is only temporary in highly disassortative networks. We explained these differences in network dynamics by a confluence of two factors: (i) frequent alternating transitions between the vaccinating and non-vaccinating phases, and (ii) scattering of vaccinator  nodes in disassortative networks. 

To evaluate the severity of epidemics, we analysed the cumulative incidence during the first outbreaks, observing a strong agreement with the results obtained for disease prevalence. We also examined the relationship between network-theoretic measures and the cumulative incidence of the first and the last wave of the epidemic, for which $r$ and $\lambda_{max}$ display a nonlinear correlation. In addition, we note that $k_s$ does not change in non-assortative and disassortative networks, and thus is inadequate for describing the relationship between network assortativity and epidemic severity.  

In summary, highly assortative scale-free networks are found to be particularly impactful in suppressing epidemics when the population follows a voluntary  vaccination scheme. In contrast, in disassortative networks, under the same conditions, the suppression only lasts for a relatively short period of time.

\section*{Declaration of interest}
The authors declare no competing interests.

\section*{Acknowledgement}
This research was funded by the Australian Research Council (ARC) Discovery Project grant DP160102742. Additionally,  S. L. C was supported by an Australian Government Research Training Program (RTP) Scholarship. Furthermore, this research was supported by the Sydney Informatics Hub at the University of Sydney, through the use of High Performance Computing (HPC) services.
\section*{Appendices}
\subsection*{Appendix A: Assortative rewiring}
\label{Appendix:A}
Assortativity here, $r$, is quantified by the Pearson correlation coefficient of the degrees at either ends of an edge ($-1 \leq r \leq 1$). For $r=0$, the network is neutral; for $r<0$, the network is disassortative; and for $r>0$, the network is assortative \cite{Newman2002,scale_free}. For scale-free networks, it has been found that the level of assortativity is constrained by the scale-free degree sequence of networks and bounded by the maximally disassortative and assortative mixing \cite{coef_limit}. A higher scale-free exponent, $\gamma$,  has a wider range of possible assortativity between $r_{min}$ (maximally disassortative) and $r_{max}$ (maximally assortative) within the scale-free regime ($2 \leq \gamma \leq 3 $) \cite{scale_free}. We, therefore, choose $\gamma=2.75$ as the scale-free exponent and use the Barab\'asi-{A}lbert model with random preferential attachment rate, $m \in [1,3]$, to construct scale-free networks from a simple 3-node fully connected network while preserving terminal nodes using \cite{igraph}.

\begin{figure*}[ht]
	\includegraphics[width=\textwidth,trim={4.5cm 1cm 1.5cm 0.5cm},clip]{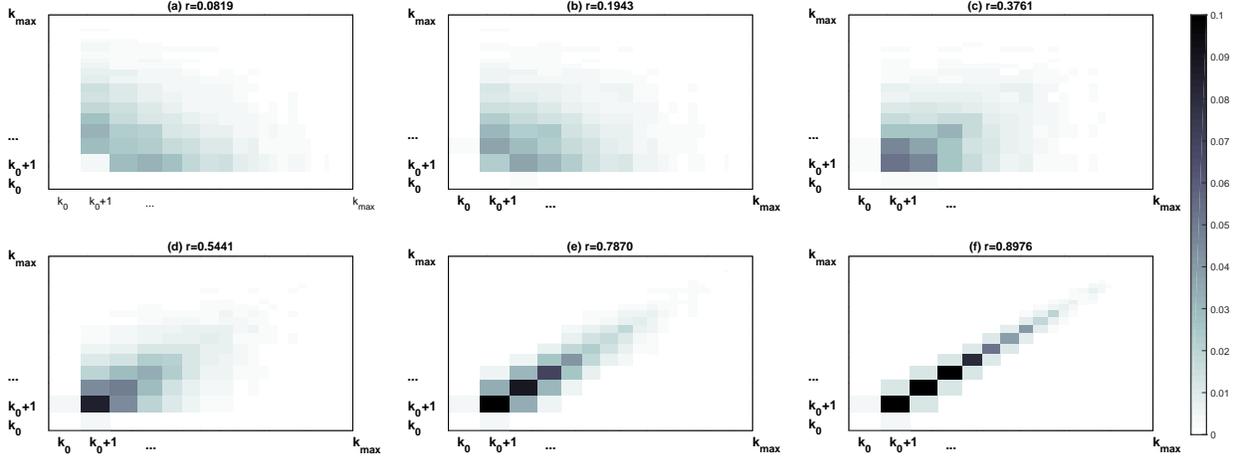}
	\caption{Assortative rewiring: adjacency matrix $\mathbf{A}$, with increasing assortativity following Xulvi-Brunet and Sokolov algorithm. $N=3000, \gamma=2.75, \langle k \rangle \approx 4,k_0=1$. Entries in $\mathbf{A}$ are ordered by increasing node degree $k$. The colour bar indicates the density of nodes in each cell. Nodes with similar degrees are connected in highly assortative networks.} 
	\label{assor_all}
\end{figure*}

\begin{figure*}[ht]
	\includegraphics[width=\textwidth,trim={4.5cm 1cm 1.1cm 0.5cm},clip]{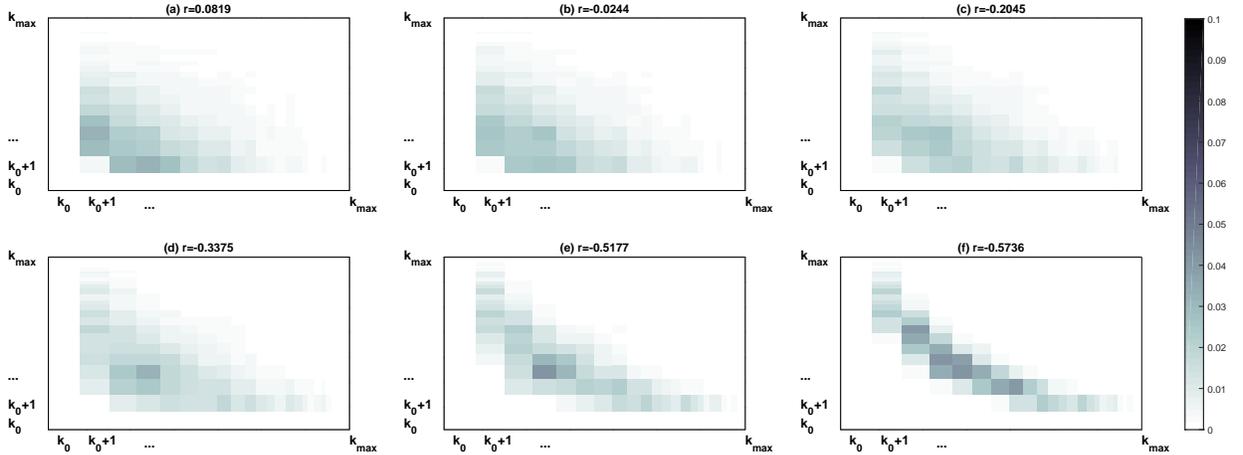}
	\caption{Disassortative rewiring: adjacency matrix $\mathbf{A}$ of networks with decreasing assortativity following Xulvi-Brunet and Sokolov algorithm. $N=3000, \gamma=2.75, \langle k \rangle \approx 4,k_0=1$. Entries in $\mathbf{A}$ are ordered to increasing node degree $k$. The color bar indicates the density of nodes in each cell. Nodes with rather different degrees are connected in highly disassortative networks.} 
	\label{disass_all}
\end{figure*}

To achieve the desired level of assortativity, we use Xulvi-Brunet and Sokolov algorithm to rewire a scale-free network while preserving its power-law degree distribution and scale-free properties \cite{Xulvi-Brunet-Sokolov}. The rewiring algorithm is as follows:
\begin{enumerate}[rightmargin=\leftmargin]
	\item Randomly select two links and locate four nodes of the selected edges. 
	\item Order the four nodes with respect to their degrees ($k$) from high to low ($a,b,c,d$ where $k_a>k_b>k_c>k_d$). 
	\item For assortative rewiring, links are formed between nodes with similar degrees (i.e., ${(a,b),(c,d)})$. For disassortative rewiring, links are formed between nodes with polarising degrees (i.e., ${(a,d),(b,c)}$).
	\item If the new links already exist in the network, the rewiring step is discarded. Go back to step 1 to select a new pair of links.
\end{enumerate} 
Clearly, higher assortativity (and disassortativity) can be achieved by higher number of rewiring. We rewire a scale-free network to achieve 15 different levels of assortativity within $-1 \leq r \leq 1$. Table \ref{assor_rewire} summarises how $r$ changes with changing number of assortative and disassortative rewiring. Figure \ref{assor_all} and \ref{disass_all} show the inner regularity of scale-free networks when nodes connect to other nodes with similar degree (Figure \ref{assor_all}) and nodes that connect to nodes with different degrees (Figure \ref{disass_all}). 
\begin{table*}
	\centering
	\begin{tabular}{c|c|c|c|c}
		\hline
		Steps & $r_a$ & $S_a$ &  $r_d$ & $S_d$ \\
		\hline
		0   & 0.0889 [0.0783,0.1053]       &0.0101       &0.0889 [0.0783,0.1053]    &0.0101\\
		700 & 0.2060 [0.1844,0.2430]      &0.0259       &-0.0282 [-0.0459,-0.0120]   &0.0366\\
		2000& 0.3891 [0.3695,0.4297]      &0.0176       &-0.1952 [-0.2218,-0.1711]   &0.0147\\
		3500& 0.5416 [0.5098,0.6114]      &0.0288       &-0.3220 [-0.2997,-0.3716]   &0.0205\\
		6000&  0.7103 [0.6827,0.7876]      &0.0306         &-0.4567 [-0.5232,-0.4201]  &0.0274 \\
		8000 & 0.7857 [0.7568,0.8592]     &0.0192       &-0.5065 [-0.5813,-0.4782]   &0.0115\\
		10000 & 0.8373 [0.8117,0.9010]     &0.0262     & -0.5354 [-0.6151,-0.5013]      & 0.0366\\
		15000 & 0.9002 [0.8733,0.9616]     &0.0292       &-0.5671 [-0.6571,-0.5281]   &0.0289\\
		\hline
	\end{tabular}
	\caption{Assortative or disassortative rewiring using Xulvi-Brunet and Sokolov algorithm. Assortativity coefficient is averaged over 10 runs. $r_a$: assortativity; $r_d$: disassortativity; bracket shows the range of $r_a$ and $r_d$ for 10 runs. $S_a$:standard deviation of $r_a$ ; $S_d$: Standard deviation of $r_d$. }
	\label{assor_rewire}
\end{table*} 

\begin{figure*}[ht]
	\includegraphics[width=\textwidth,trim={1cm 0.5cm 1.5cm 0.5cm},clip]{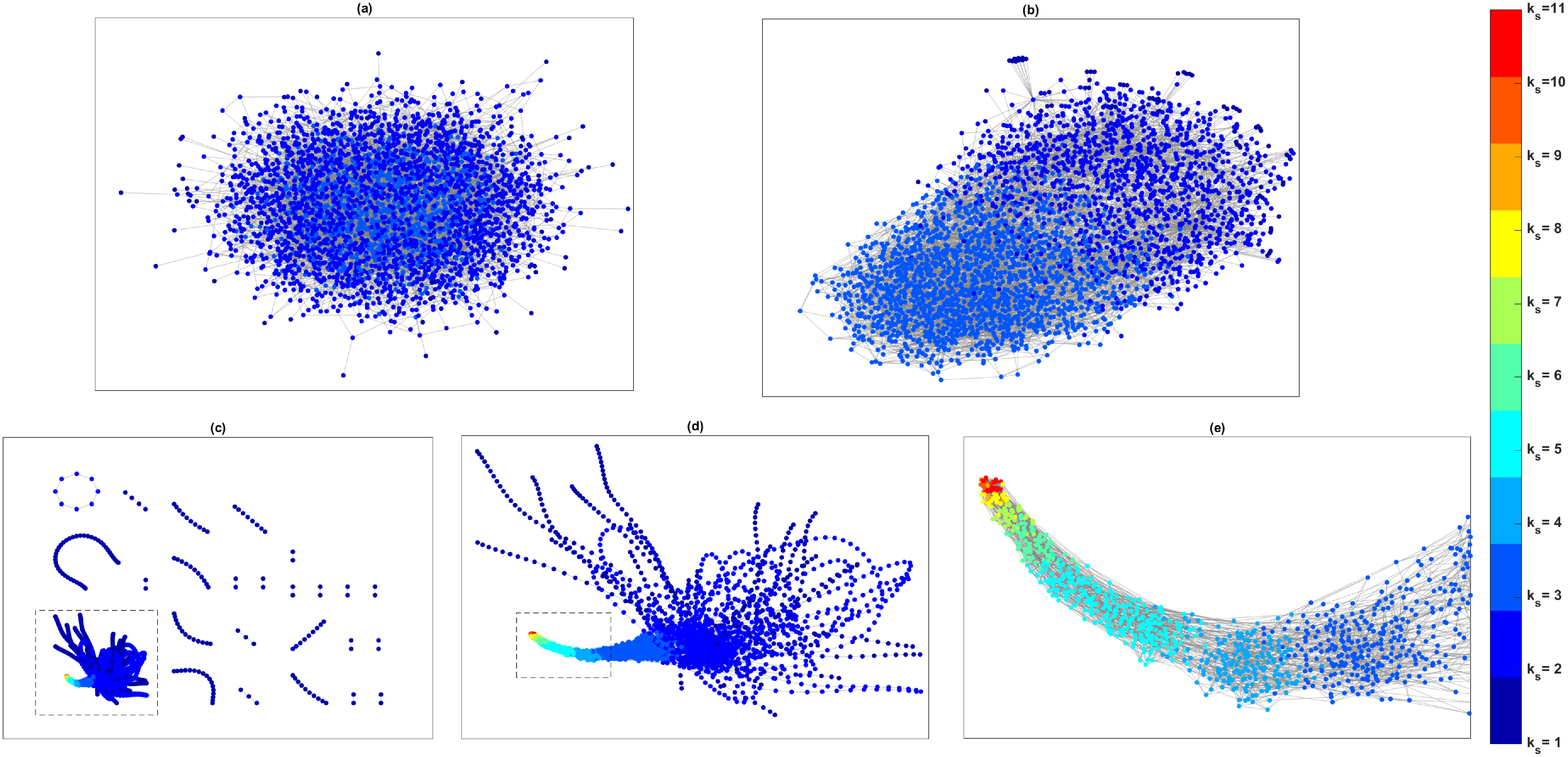}
	\caption{Maximum $k_s$ increases with assortative rewiring. (a) original network $r=0.0819$, (b) highly disassortative network $r=\text{-}0.5736$, (c)-(e) assortative rewiring $r=0.8976$. (d) and (e) are magnifying sections of (c) showing the high coreness found in assortative networks. Other network properties: $N=3000, \gamma=2.75, \langle k \rangle \approx 4,k_0=1$.} 
	\label{rewire_k_s}
\end{figure*}

Assortative rewiring also affects the maximum coreness $k_s$. With assortative rewiring, while low degree nodes are unaffected, hubs are connected to other hubs and thus become less dependent on the low-degree nodes. The connection between hubs form a more cohesive sub-network at higher $k\text{-}core$ \cite{Newman2002}. Disassortative rewiring, on the other hand, makes hubs more dependent on low-degree nodes so that the network should have fewer shells. We illustrate this dependency between assortativity ($r$) and maximum coreness ($k_s$) in Figure \ref{rewire_k_s}. \\

\subsection*{Appendix B: Epidemic parameters and cumulative incidence}
\label{Appendix:B}
To further assess the risk of measles, we compute the cumulative incidence ($CI$) during the first outbreak. Cumulative incidence quantifies the proportion of population at risk, defined as the ratio between the number of new cases divided by the total population at risk over a specific period of time. Disease prevalence ($I$), on the other hand, is measured at one point in time \cite{epi_book}. These two measures are closely related and the cumulative incidence during the first outbreak $T$ is determined as:

\begin{equation}
CI(T)=\sum\limits_{t=1}^{T}\sum\limits_{j=1}^M \sum\limits_{k=1}^{M} \beta_j \phi_{ij} \frac{\phi_{kj}I_k}{\epsilon ^p_j} S_i
\end{equation}

Where $T$ is measured from initial conditions $t=0, I_0=0.001$ and to the point where the prevalence is below the threshold value $ I < \hat{I}$ where $\hat{I}=0.0002$.

Unless stated otherwise, the behavioural and epidemiological parameters used in simulations are summarised in Table \ref{parameter}.

\begin{table*}[tbh!]
	\renewcommand{\arraystretch}{1.2}
	\begin{tabular}{|c |c|c| c|}
		\hline
		Parameter  & Interpretation                                  & Baseline value & References \\
		\hline 
		$\beta$    & Transmission rate ($day^{-1}$)                  & 1.5              &             \\
		$1/\gamma$ & Average length of recovery period (days)        & 10             & \cite{measle_R0}    \\
		$R_0$      & Basic reproduction number                       & 15             & \cite{measle_R0}\\ 
		$\mu$      & Mean birth and death rate                       & 0.000055       &             \\
		$\kappa$   & Imitation rate                                  & 0.001          & \cite{bauch_2005_imi}      \\
		$\omega$   & Responsiveness to changes in disease prevalence & 3500    & \cite{bauch_2005_imi}     \\
		$\phi_{ij}$& Fraction of residents from node $i$ travelling to $j$ & [0,1]    & Network connectivity \\
		\hline   
	\end{tabular}
	\caption{Epidemiological and behavioural parameters used in simulation.}
	\label{parameter}
\end{table*}

\subsection*{Appendix C: Partitioned network dynamics}
\label{Appendix:C}

\begin{figure*}
	\centering
	\includegraphics[width=\textwidth,trim={3cm 1cm 3cm 1cm},clip]{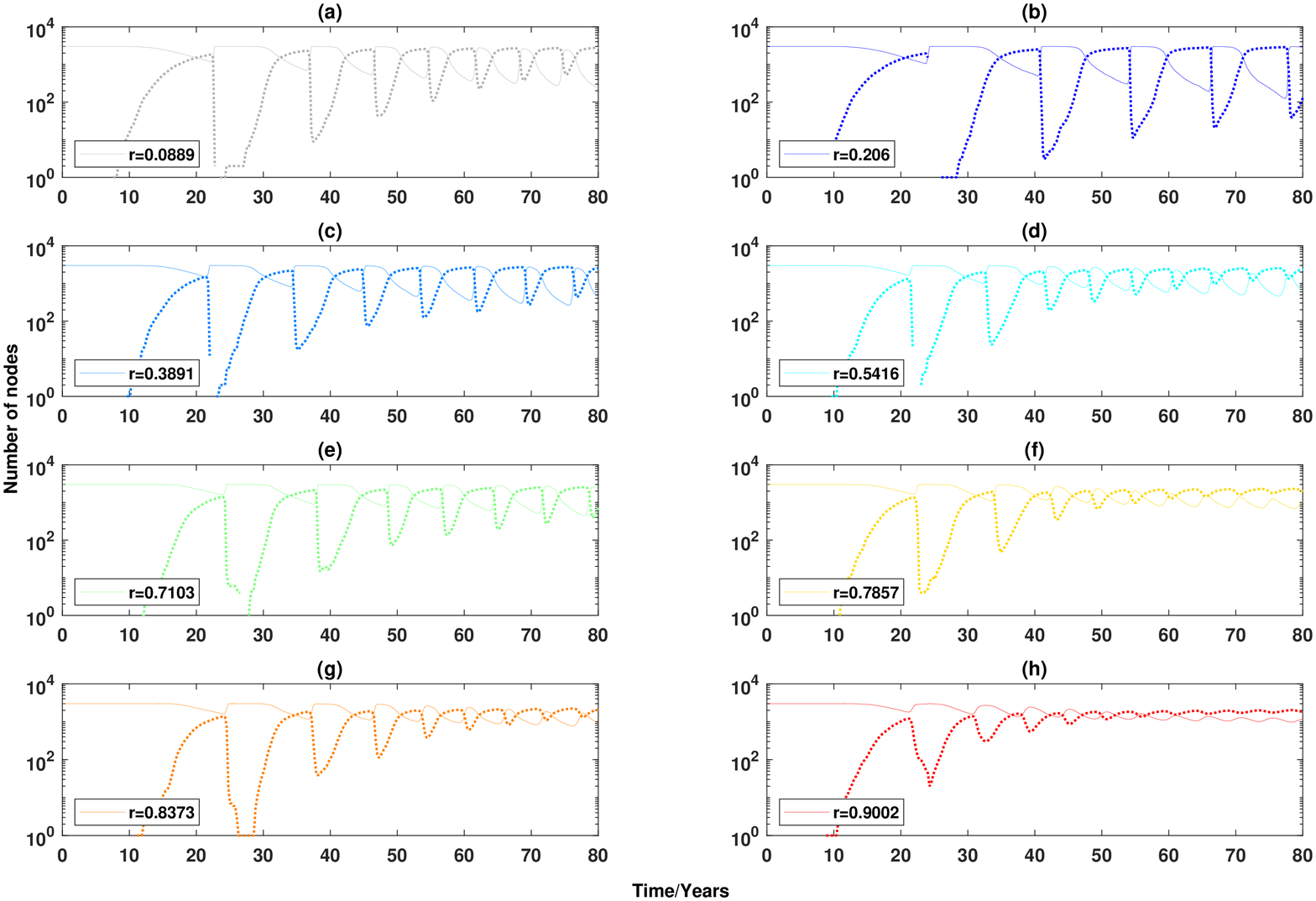}
	\caption{Number of nodes in the sub-networks formed by majority vaccinators  and majority non-vaccinators, traced over time in the assortative range. Solid line: majority vaccinators; dashed line: majority non-vaccinators. We refer to Figure \ref{osci_ass_3500} for the setting of epidemic and vaccination dynamics and parameters used.} 
	\label{node_ass}
\end{figure*}

\begin{figure*}
	\centering
	\includegraphics[width=\textwidth,trim={3cm 1cm 3cm 1cm},clip]{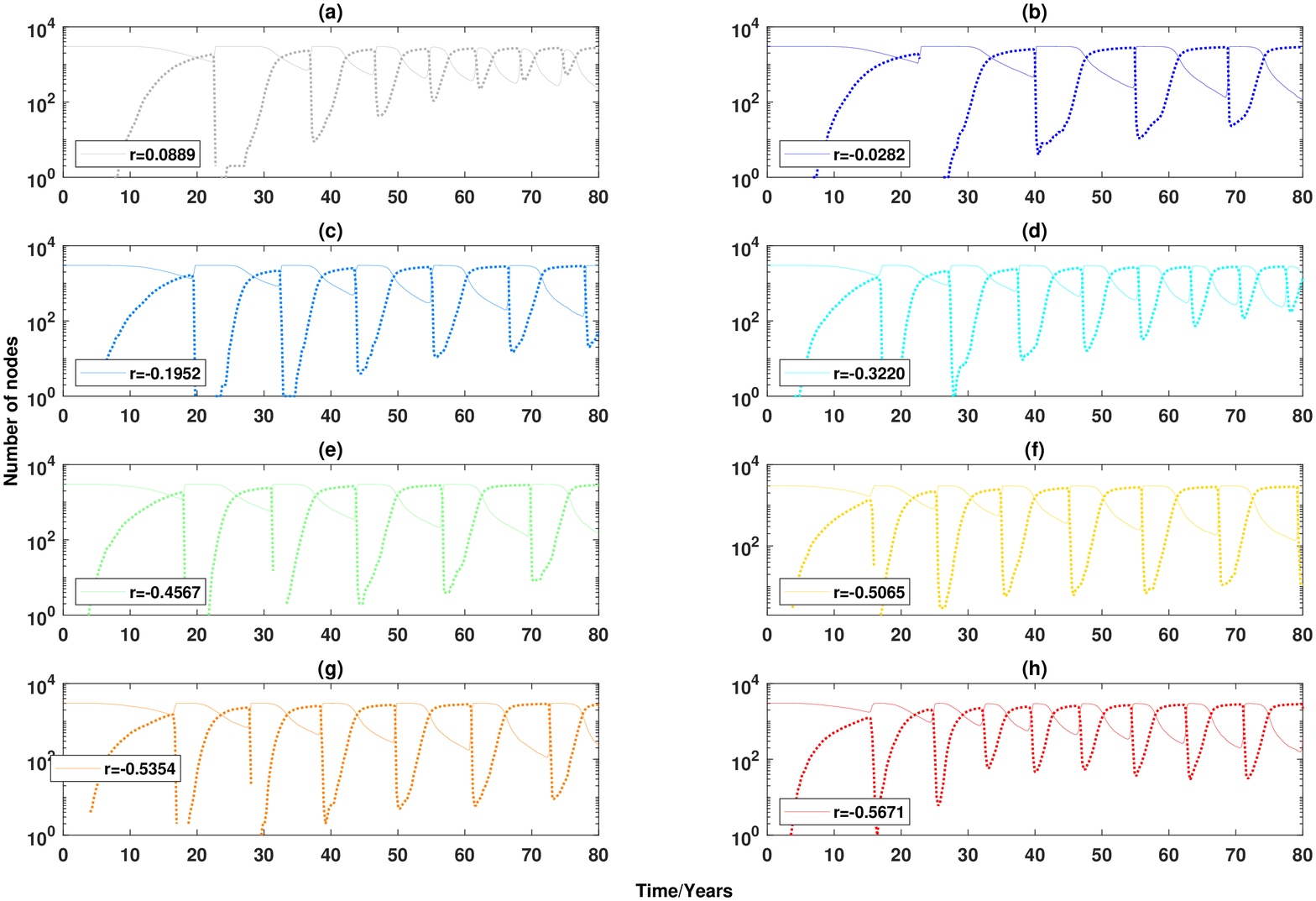}
	\caption{Number of nodes in the sub-networks formed by majority vaccinators  and majority non-vaccinators, traced over time in the disassortative range. Solid line: majority vaccinators; dashed line: majority non-vaccinators. We refer to Figure \ref{osci_dis_3500} for the setting of epidemic and vaccination dynamics and parameters used.} 
	\label{node_dis}
\end{figure*}

\begin{figure*}
	\centering
	\includegraphics[width=\textwidth,trim={3cm 1cm 3cm 1cm},clip]{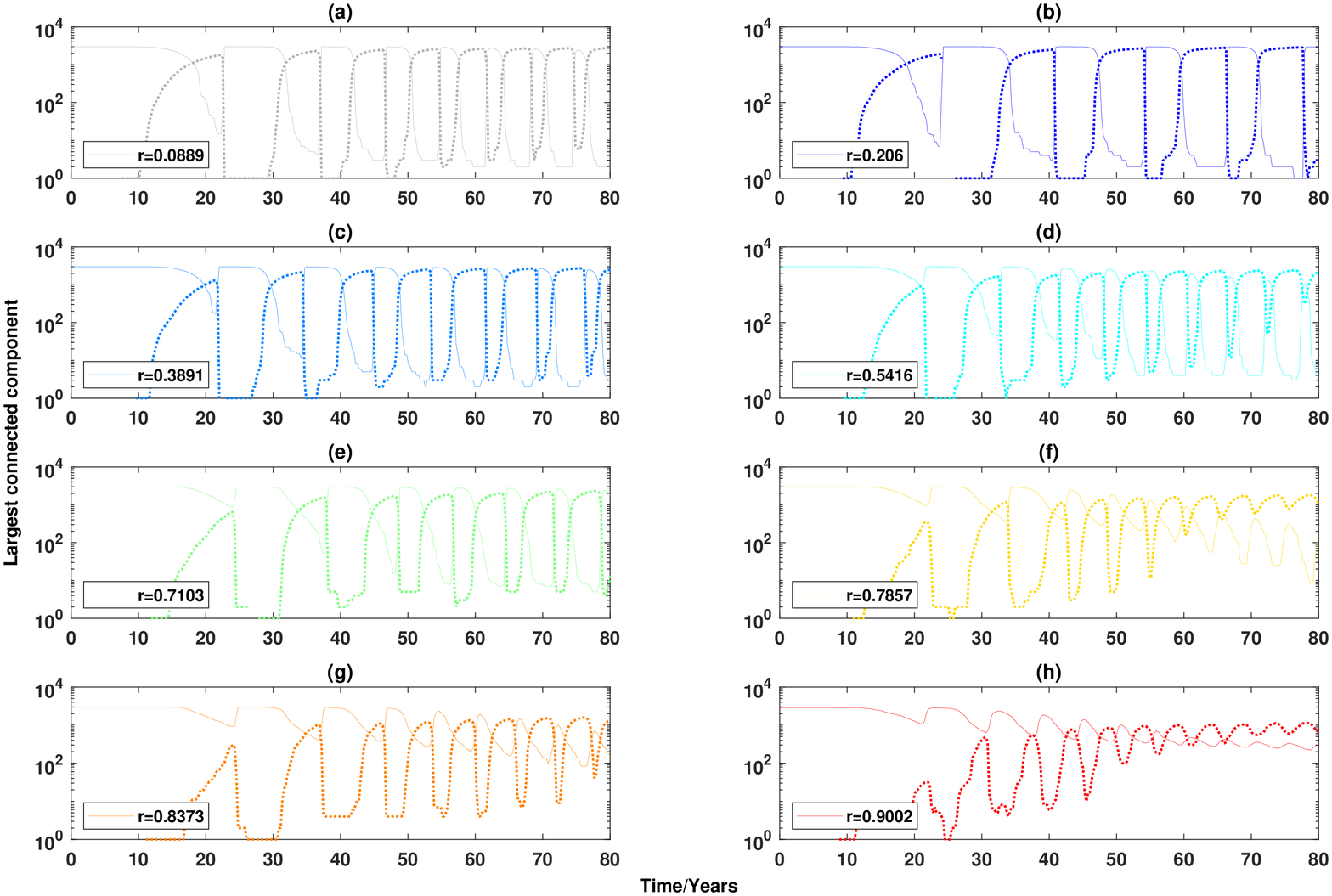}
	\caption{Largest connected subgraph in the sub-networks formed by majority vaccinators  and majority non-vaccinators, traced over time in the assortative range. Solid line: majority vaccinators; dashed line: majority non-vaccinators. We refer to Figure \ref{osci_ass_3500} for the setting of epidemic and vaccination dynamics and parameters used.} 
	\label{LCS_ass}
\end{figure*}

\begin{figure*}
	\centering
	\includegraphics[width=\textwidth,trim={3cm 1cm 3cm 1cm},clip]{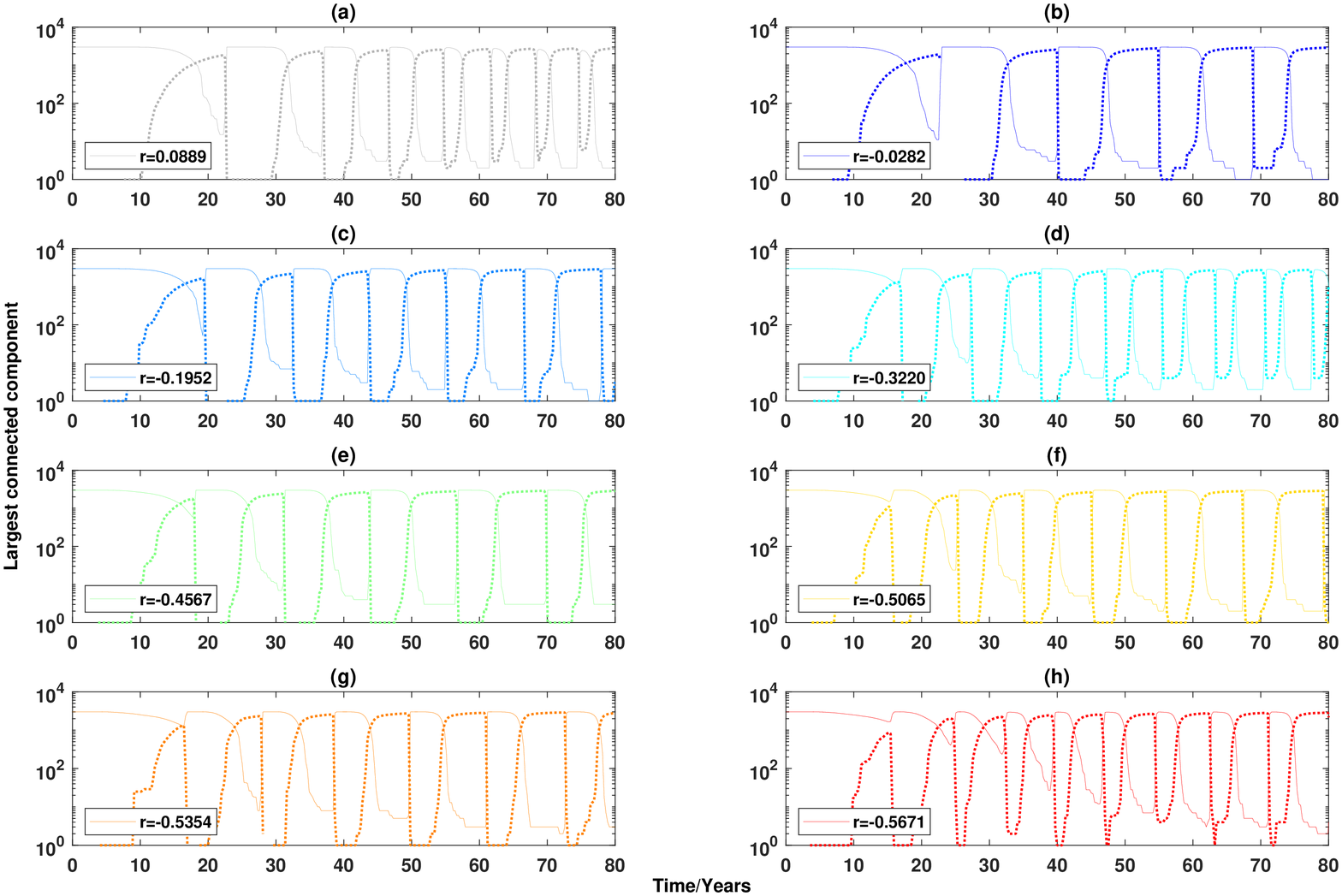}
	\caption{Largest connected subgraph in the sub-networks formed by majority vaccinators  and majority non-vaccinators, traced over time in the disassortative range. Solid line: majority vaccinators; dashed line: majority non-vaccinators. We refer to Figure \ref{osci_dis_3500} for the setting of epidemic and vaccination dynamics and parameters used.} 
	\label{LCS_dis}
\end{figure*}

In this appendix we explore network dynamics within the partitioned majority networks. This analysis is carried out for assortative and disassortative ranges, in terms of (i) the number of nodes, contrasted  in Figures \ref{node_ass} and \ref{node_dis} respectively, and (ii) the sizes of largest connected subgraphs (LCS), compared  in Figures \ref{LCS_ass} and \ref{LCS_dis}.

\bibliographystyle{unsrt}
\bibliography{ref_2019}
\end{document}